\documentclass[a4paper,fleqn]{cas-dc}

\usepackage{stfloats}
\makeatletter
\newcommand{\dblbottom}{\def\@dbflt##1{\@ifnextchar[{\@xdblfloat{##1}}{\@xdblfloat{##1}[b]}}}
\newcommand{\dbltop}{\def\@dbflt##1{\@ifnextchar[{\@xdblfloat{##1}}{\@xdblfloat{##1}[tbp]}}}
\makeatother

\usepackage[numbers]{natbib}

\usepackage{threeparttable}

\usepackage{tabularx}
\newcolumntype{L}{>{\raggedright\arraybackslash}X}

\begin{document}
\let\WriteBookmarks\relax
\def\floatpagepagefraction{1}
\def\textpagefraction{.001}

\shorttitle{CREMA setup at the EMA Laboratory, MIT}
\shortauthors{F.~C.~Pastrana Cruz et~al.}

\title[mode = title]{\textsc{CREMA}: Collinear Resonance for Exotic Molecules and Atoms}

\author[1]{Fabian~C.~Pastrana~Cruz}[orcid=0009-0008-7469-7513]
\cormark[1]
\ead{pastrana@mit.edu}

\author[1]{Shane~G.~Wilkins}[orcid=0000-0001-8897-7227]
\fnmark[1]

\author[1]{Alex~J.~Brinson}
\author[2]{Paul~Fischer}
\author[1]{Mathieu~Flayol}
\author[3,1]{Derick~Gonzalez-Acevedo}
\author[1]{Jackson~Hacias}
\author[4]{Jason~Huskey}  
\author[1]{Haruka~Kakioka}
\author[5]{Carina~Kanitz}
\author[1]{Jonas~Karthein}[orcid=0000-0002-4306-9708]
\fnmark[2]
\author[5]{Simon~Lechner}
\author[5,6,7]{Stephan~Malbrunot-Ettenauer}
\author[1]{Scott~Moroch}
\author[1]{Jose~M.~Munoz}
\author[8]{Moritz~P.~Reiter}
\author[9,10]{Mikael~Reponen}
\author[2]{Lutz~Schweikhard}[orcid=0009-0002-8272-0388]
\author[1,11,12]{Silviu~M.~Udrescu}
\author[1]{Adam~R.~Vernon}
\author[5]{Markus~Vilen}
\author[1,13]{Ronald~F.~Garcia~Ruiz}
\cormark[1]
\ead{rgarciar@mit.edu}

\affiliation[1]{organization={Laboratory for Nuclear Science, Massachusetts Institute of Technology},
                city={Cambridge},
                state={Massachusetts},
                postcode={02139},
                country={USA}}

\affiliation[2]{organization={Institute of Physics, University of Greifswald},
                postcode={17487},
                city={Greifswald},
                country={Germany}}

\affiliation[3]{organization={Department of Physics, Harvard University},
                city={Cambridge},
                state={Massachusetts},
                postcode={02138},
                country={USA}}
\affiliation[4]{organization={Department of Physics, University of California},
                city={Berkeley}, 
                state={California},
                postcode={94720}, 
                country={USA}}

\affiliation[5]{organization={Experimental Physics Department, CERN},
                postcode={CH-1211},
                city={Geneva 23},
                country={Switzerland}}

\affiliation[6]{organization={TRIUMF},
                addressline={4004 Wesbrook Mall},
                city={Vancouver},
                state={British Columbia},
                postcode={V6T 2A3},
                country={Canada}}

\affiliation[7]{organization={Department of Physics, University of Toronto},
                city={Toronto},
                state={Ontario},
                postcode={M5S 1A7},
                country={Canada}}
                
\affiliation[8]{organization={School of Physics \& Astronomy, The University of
                Edinburgh},
                addressline={Peter Guthrie Tait Road},
                postcode={EH9 3FD},
                city={Edinburgh},
                country={United Kingdom}}        
\affiliation[9]{organization={University of Jyv\"askyl\"a},
                addressline={P.O. Box 35},
                postcode={FI-40014},
                city={Jyv\"askyl\"a},
                country={Finland}}
\affiliation[10]{organization={Helsinki Institute of Physics},
                addressline={P.O. Box 64},
                postcode={FI-00014},
                city={Helsinki},
                country={Finland}}

\affiliation[11]{organization={Department of Physics, Princeton University},
                city={Princeton},
                state={New Jersey},
                postcode={08544},
                country={USA}}
\affiliation[12]{organization={Department of Physics and Astronomy, Johns Hopkins University},
                city={Baltimore},
                state={Maryland},
                postcode={21210},
                country={USA}
                }
    
\affiliation[13]{organization={Harvard-MIT Center for Ultracold Atoms},
                city={Cambridge},
                state={Massachusetts},
                postcode={02138},
                country={USA}}

\cortext[1]{Corresponding authors.}
\fntext[1]{Present address: Facility for Rare Isotope Beams, Michigan State
           University, East Lansing, USA}
\fntext[2]{Present address: Cyclotron Institute \& Department of Physics and Astronomy, Texas A\&M University, College Station, Texas 77840, USA}

\begin{abstract}
The Collinear Resonance for Exotic Molecules and Atoms (CREMA) setup has been developed and commissioned at the Massachusetts Institute of Technology's Laboratory for Nuclear Science. The instrument combines the Doppler-suppressed resolution of the collinear fast-beam geometry with the selectivity of resonance ionization, for precision measurements of hyperfine structure and isotope shifts in atoms and molecules. We describe the setup, including the ion-optical simulations in which a multi-objective evolutionary algorithm provided the starting settings for the beam transport, and report commissioning measurements of the hyperfine structure of $^{133}$Cs. The extracted hyperfine constants agree with literature values within one standard deviation, demonstrating the operational readiness of the setup. CREMA mirrors the Resonance Ionization Spectroscopy Experiment at the Facility for Rare Isotope Beams and serves as its offline counterpart for the development of laser ionization schemes and techniques for future studies of exotic atoms and molecules.
\end{abstract}

\begin{keywords}
collinear laser spectroscopy \sep resonance ionization \sep hyperfine structure
\sep ion-optical simulations
\end{keywords}

\maketitle

\section{\label{sec:level1}Introduction }
Laser spectroscopy has proven to be a powerful tool for the study of nuclear electromagnetic properties of ground and long-lived isomeric states. Through hyperfine structure and isotope shift measurements, the nuclear spin, nuclear magnetic dipole and electric quadrupole moments, and changes in the nuclear mean-square charge radii can be precisely extracted in a nuclear model-independent way. Measuring these fundamental nuclear properties remains critical to our understanding of nuclei and the microscopic description of the nuclear force \cite{Yang2023}. 

Among the available techniques, collinear resonance ionization spectroscopy (CRIS)~\cite{Kudriavtsev1982,Sch91} is particularly well suited to the most exotic species, combining the high resolution of collinear fast-beam spectroscopy with the high selectivity and efficiency of multi-step resonance ionization. Pioneered at ISOLDE~\cite{Sch91,Fla13,Cocolios2013,Coc16}, CRIS has enabled access to a wide range of exotic isotopes, guiding our understanding of atomic nuclei far from stability~\cite{Vernon2022,Koszorus2021,deGroote2020,Karthein2024}. More recently, CRIS has been extended to short-lived radioactive molecules~\cite{GarciaRuiz2020,Udrescu2024RaF,Wilk25,Ath25}, which are of interest for tests of fundamental symmetries. The technique has since been implemented at different laboratories worldwide~\cite{Kuj23,Bri26,Mei2026,Hu2025}, including the recently commissioned Resonance Ionization Spectroscopy Experiment (RISE)~\cite{Bri26} at the BEam COoler and LAser spectroscopy (BECOLA) facility~\cite{Minamisono2013CommissioningNSCL} at the Facility for Rare Isotope Beams (FRIB), whose online campaign enabled the first laser-spectroscopy measurements of the charge radii of neutron-deficient aluminum isotopes out to the proton drip line~\cite{BrinsonAl2026}.

Successful experimental campaigns with these online instruments, however, require complementary offline infrastructure. Laser spectroscopy measurements of previously unstudied atomic properties typically require developing dedicated multi-step ionization schemes, identifying suitable transitions, setting up new laser systems, optimizing many parameters, and characterizing systematic uncertainties. This preparatory work can take months, which is impractical to sustain at an online facility where beam time is allocated competitively and shared across a broad user community. 
\dblbottom
\begin{figure*}[pos=!htb]
\centering
\includegraphics[width=1\textwidth]{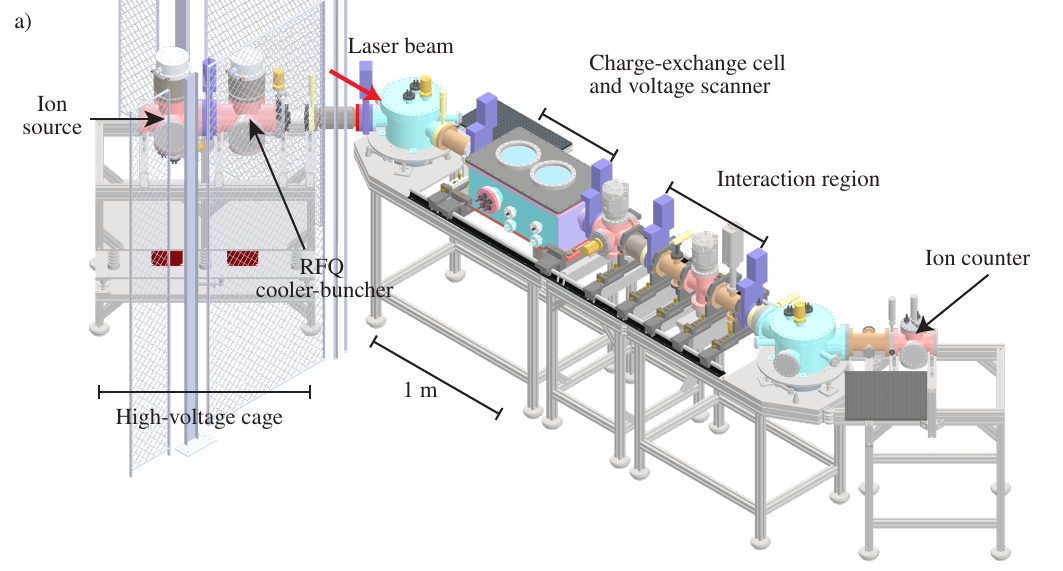}

\medskip

\includegraphics[width=1\textwidth]{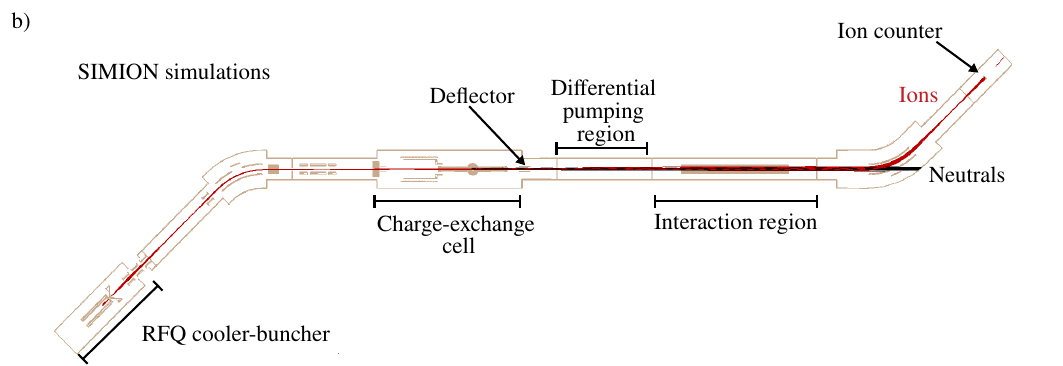}

\caption{(a) CAD diagram of the CREMA setup. Ions from the surface ionization source are cooled and bunched in the RFQ cooler-buncher inside the high-voltage cage, accelerated through 10 or 15~kV into the grounded beamline, neutralized in the charge-exchange cell (CEC), and overlapped collinearly with the laser beams (red arrow) in the interaction region. The resonantly ionized atoms are deflected onto the ion counter.
(b) SIMION simulation from the exit of the RFQ cooler-buncher to the ion counter, with ions in red and neutral atoms in black. At the CEC, some of the simulated ions are neutralized and travel straight through the interaction region. The deflector after the CEC is off in the simulation so that both paths are shown. In operation, it removes the remaining ions. At the final bender, the ion beam is bent onto the counter (``Ions'') and the neutral atoms continue straight (``Neutrals'').}
\label{fig:experimental_layout}

\end{figure*}
\dbltop
Motivated by these challenges, we have constructed and commissioned the Collinear Resonance for Exotic Molecules and Atoms (CREMA) setup at the Exotic Molecules and Atoms (EMA) Laboratory at the Massachusetts Institute of Technology (MIT). CREMA mirrors the architecture of RISE, so that schemes and tools developed at MIT can be transferred to FRIB, while remaining flexible enough to support studies of radioactive molecules and the development of new techniques targeting future campaigns with exotic isotopes. In this work, we describe the design of the apparatus and report on its first commissioning measurements on the $6s\,^2S_{1/2}\to 8p\,^2P_{1/2}$ and $6s\,^2S_{1/2}\to 8p\,^2P_{3/2}$ transitions of stable $^{133}$Cs (labeled below by their upper states, $^2P_{1/2}$ and $^2P_{3/2}$), which demonstrate operational readiness and quantify the current resolution and stability of the system.

\section{Ion beamline}
The layout of CREMA is shown in Fig.~\ref{fig:experimental_layout}. The setup has three main parts: (i) the ion source and the radio-frequency quadrupole (RFQ) cooler-buncher, both on a high-voltage platform enclosed by the high-voltage cage; (ii) the charge-exchange cell (CEC); and (iii) the interaction region and ion detection.

\subsection{Ion source and RFQ cooler-buncher}
\begin{figure}[pos=!ht]
    \centering
    \includegraphics[width=1\linewidth]{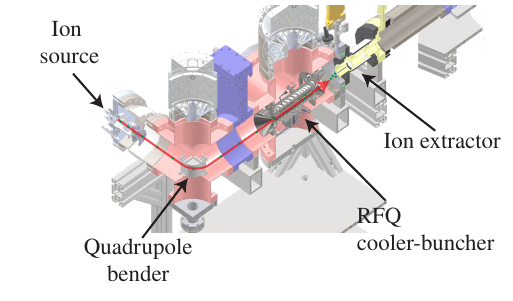}
    \caption{Cross-sectional view of the beamline on the high-voltage platform. The red line traces the ion beam from the surface ionization cesium source through the quadrupole bender into the RFQ cooler-buncher. The ion extractor at its exit accelerates the bunches from the platform potential into the grounded beamline.}
    \label{fig:HVC-Vacuum}
\end{figure}

For commissioning, a surface ionization cesium source was installed on the high-voltage platform. Surface ionization sources provide continuous beams of alkali ions with an energy spread below 1~eV~\cite{Flowerdew2021}. The beamline on the high-voltage platform is shown in Fig.~\ref{fig:HVC-Vacuum}. Singly charged ions are produced and accelerated to 125~eV with respect to the high-voltage platform, deflected by an electrostatic quadrupole bender, and transported to a gas-filled linear RFQ cooler-buncher, a digitally operated trap developed for the MIRACLS project~\cite{Maier2023,SELS2020} and described in Refs.~\cite{Carina21,Croquette2023,Giesel2024}. The RFQ cooler-buncher traps the ions and cools them through collisions with an argon buffer gas. Once cooled, the ions can be released in bunches, which in the present measurements were 1.2 to 1.6~$\mu$s long (FWHM).

After the RFQ cooler-buncher, bunches of ions are accelerated by a potential difference of 10 or 15~kV between the high-voltage platform and the rest of the beamline at ground potential. Subsequently, the ion beam is bent through $45^\circ$ toward a quadrupole triplet and steering electrodes, which are used to tune the beam through the CEC.

The $45^\circ$ bend moves the ion source and the RFQ cooler-buncher off the laser axis, so that the laser beams can be sent along the full length of the interaction region while optical access is kept for alignment and diagnostics. The transport is tuned section by section, starting from the settings found in the simulations of Sec.~\ref{section:simulation}.

\subsection{Charge-exchange cell}
\begin{figure}[pos=!ht]
    \includegraphics[width=.95\linewidth]{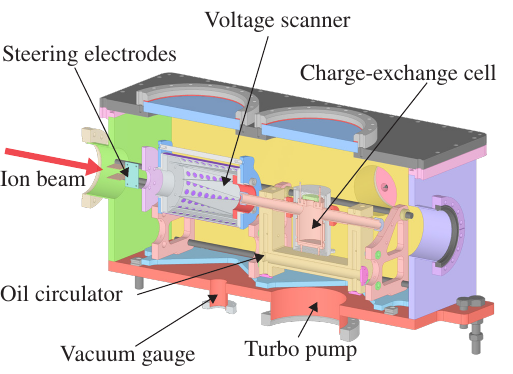}
    \caption{Cross-sectional view of the CEC chamber. The ion beam enters from the left through a set of steering electrodes and passes through the voltage scanner electrodes before reaching the alkali-filled CEC where neutralization takes place. The chamber is pumped down by a turbomolecular pump monitored with a vacuum gauge, and the oil circulator maintains the temperature of the CEC end caps.}
    \label{fig:CEC_cross-section}
\end{figure}

Ions pass through a CEC, where they interact with an alkali vapor, used for in-flight neutralization of the ion beam. The bunches then proceed into a differential pumping region, where additionally an electric field is applied to deflect away non-neutralized components of the beam. A cross-sectional view of the CEC chamber is shown in Fig.~\ref{fig:CEC_cross-section}. The design of the CEC chamber is based on the setup developed in Refs.~\cite{Mei2026,Gins2019,Wang2022,Bai2022}. 

Both sodium and potassium were used as neutralization media during commissioning, both predicted to populate mainly the ground state of the Cs atom at a beam energy of 10~keV~\cite{VERNON201961,DOCKERY2025}. For sodium, the reservoir was held at 380~$^{\circ}$C and the end caps at 110~$^{\circ}$C. For potassium, the reservoir was kept at 300~$^{\circ}$C and the end caps at 95~$^{\circ}$C. 

The cell is mounted on an insulated support so that it can be biased relative to the beamline, which changes the beam energy in the neutralization region and, with it, the Doppler shift (Doppler tuning). The high-voltage hardware for this voltage scanner was not yet installed, and all spectra in this work were recorded by scanning the laser frequency at fixed beam energy.

\subsection{Interaction region and ion detection}
After the differential pumping region, the neutral bunches enter the interaction region, about 1~m long, where they are overlapped in space and time with pulsed laser beams that co-propagate with them. One or more lasers resonantly excite the atoms or molecules to a chosen state, from which a high-power pulsed laser ionizes them non-resonantly.

To suppress possible background from non-resonant collisional ionization, the interaction region was kept under ultra-high vacuum ($\leq 10^{-9}$~mbar). The differential pumping region between the CEC and the interaction region limits the migration of alkali vapor into this section. 

Following resonant excitation and ionization, the ions are deflected by the final bender onto the ion counter, a MagneTOF detector (ETP Ion Detect) operated in single-ion counting mode. Because the beam is bunched, only ions arriving within a time-of-flight gate of 3 to 4~$\mu$s around the expected bunch arrival are used in the analysis, which rejects the counts that are not correlated with the ion bunches.

\section{Laser systems}\label{sec:lasers}

\begin{figure*}[pos=!t]
    \includegraphics[width=1\linewidth]{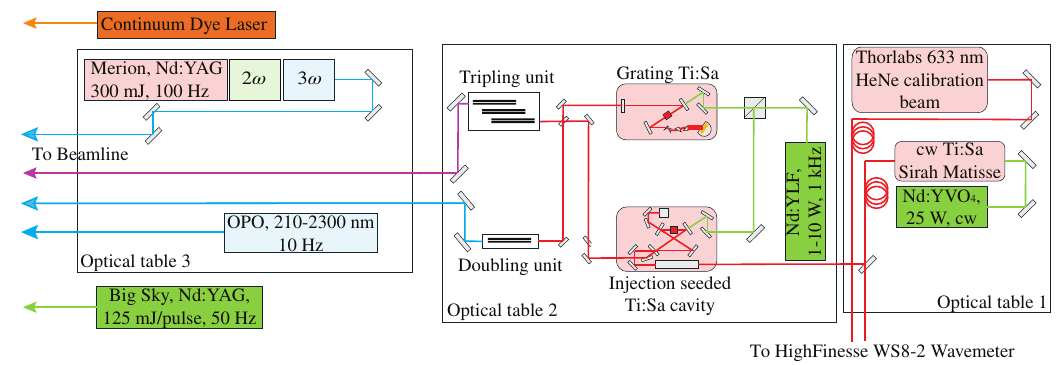}
    \vspace{-.5 cm}
    \caption{Schematic of the CREMA laser system. Box and line colors indicate the wavelength: red (633--1064~nm), green (527--532~nm), blue (350--500~nm), violet ($\lesssim$350~nm) and orange (dye laser). Arrows mark the beams sent to the beamline and to the wavemeter. The lasers are described in Section~\ref{sec:lasers} and in Table~\ref{tab:laser_summary}.}
    \label{fig:laser_setup}
\end{figure*}

A schematic of the laser setup is shown in Fig.~\ref{fig:laser_setup}, and a summary of the systems, with their main characteristics, is presented in Table~\ref{tab:laser_summary}. All lasers are housed in an independent laboratory space with temperature and humidity control, designed to maintain temperature variations below 1~$^\circ$C and a relative humidity of approximately 50\%.

\subsection{Resonant excitation laser for spectroscopy}

The laser light for resonant excitation is provided by an injection-seeded Ti:Sapphire laser system~\cite{ref:reponen_tisa}. A bow-tie Ti:Sapphire cavity is pumped at a 1~kHz repetition rate by a Nd:YLF laser (Continuum Terra, 527~nm) with $\sim$1~mJ pulse energy and a 100--150~ns pulse duration. A continuous-wave (cw) Ti:Sapphire laser (Sirah Matisse CS or M Squared SolsTiS) seeds the cavity through a single-mode optical fiber, and the gain-switched cavity emits pulses of $\sim$200~$\mu$J and $\sim$80--100~ns (FWHM), with a linewidth of $\sim$15~MHz reported for a similar laser system~\cite{sonnenschein2015thesis}. The frequency is scanned by tuning the seed laser. An active lock keeps the pulsed cavity resonant with the seed, so that its output follows.

The frequency of the cw seed laser is monitored with a HighFinesse WS8-2 wavemeter, which is periodically calibrated against a frequency-stabilized helium--neon (HeNe) laser at 633~nm.

Additional lasers, not used in the present measurements, support transition searches and extend the wavelength range beyond that of Ti:Sapphire (Table~\ref{tab:laser_summary}). A second Ti:Sapphire laser, tuned by a grating~\cite{Naubereit2014,Teigelhofer2010} and pumped by the same Nd:YLF laser, scans continuously over $\sim$700--1000~nm with up to 200~$\mu$J per pulse and a linewidth of $\sim$3~GHz in the fundamental. This is too broad for most hyperfine-structure measurements, but the wide tuning range is useful to search for transitions that have not been measured before and to scan Rydberg and autoionizing states. A transition found this way is then measured at high resolution with the injection-seeded laser.

A Quantum Light Instruments Q-TUNE-F10-SH optical parametric oscillator (OPO) covers 410--2300~nm, and 210--410~nm with its second-harmonic extension, with pulses shorter than 5~ns, linewidths below 180~GHz and up to 9~mJ per pulse in the visible~\cite{QTuneManual}. The OPO is used for transition searches in species with poorly known level structures. A Continuum ND6000 pulsed dye laser, pumped by a Nd:YAG laser, provides tunable visible light with interchangeable dyes, extended into the ultraviolet by harmonic generation. Its linewidth is below 0.08~cm$^{-1}$ (about 2.4~GHz) at 560~nm.

\subsection{Ionization laser}

The non-resonant ionization step is provided by one of two pulsed Nd:YAG lasers. A Quantel Merion MW G7-100, used for the present measurements, delivers 300~mJ at its fundamental wavelength of 1064~nm in 5--9~ns (FWHM) pulses at up to 100~Hz. Its doubling and tripling modules serve schemes in which a single 1064~nm photon cannot ionize the excited atoms or molecules. A Big Sky EMAS laser delivers up to 125~mJ at 532~nm in 10~ns pulses at up to 56~Hz.

\begin{table*}[pos=t]
\caption{Summary of laser systems available at the CREMA setup. Values are approximate and configuration dependent.}
\centering
\footnotesize
\setlength{\tabcolsep}{4pt}
\renewcommand{\arraystretch}{1.2}

\begin{tabularx}{\textwidth}{>{\raggedright\arraybackslash}p{2.5cm} L >{\raggedright\arraybackslash}p{1.4cm} L L L}
\hline\hline
System & Wavelength range & Pulse width & Pulse energy & Linewidth & Primary use \\
\hline

Injection-seeded Ti:Sapphire &
$\sim$700--1000~nm; UV via harmonic generation &
ns-scale &
$\sim$0.2~mJ (fundamental, 1~kHz) &
$\sim$15~MHz &
High-resolution excitation \\

Grating-tuned Ti:Sapphire &
$\sim$700--1000~nm; UV via harmonic generation &
ns-scale &
$\sim$0.2~mJ (fundamental, 1~kHz) &
$\sim$3~GHz &
Broadband scans and transition searches \\

Q-TUNE-F10-SH OPO &
210--2300~nm &
$<5$~ns &
up to $\sim$9~mJ (signal); $\lesssim$1~mJ (UV, idler) &
$<180$~GHz &
Transition searches and exploratory spectroscopy \\

Continuum ND6000 dye laser &
372--722~nm pumped with 355~nm; 541--900~nm pumped with 532~nm &
ns-scale &
up to tens of mJ (pump and dye dependent) &
$<0.08$~cm$^{-1}$ ($\approx$2.4~GHz) at 560~nm (Rhodamine~590) &
Alternative resonant excitation pathways \\

Nd:YAG (Merion MW / Big Sky) &
1064, 532, 355~nm &
5--10~ns &
300~mJ at 1064~nm (Merion MW); $\lesssim$125~mJ at 532~nm (Big Sky) &
GHz-scale &
Non-resonant ionization and pumping \\

\hline\hline
\end{tabularx}
\label{tab:laser_summary}
\end{table*}

\section{Experimental control and data acquisition}
\subsection{Timing and synchronization}
The laser systems are synchronized with the bunched ion beam produced by the RFQ cooler-buncher using a Quantum Composer 9520 series digital pulse generator. The RFQ cooler-buncher is usually operated at 50 or 100~Hz, depending on the repetition rate of the ionization laser in use (50~Hz for the Big Sky, 100~Hz for the Merion MW). For the present measurements, the Merion MW at 1064~nm was used at 100~Hz. The RFQ cooler-buncher released a bunch with every pulse of the ionization laser in the $^2P_{1/2}$ measurement, and with every fourth pulse, at 25~Hz, in the $^2P_{3/2}$ measurement.

Both the injection-seeded Ti:Sapphire excitation laser and the Nd:YAG ionization laser are triggered so as to coincide with each atomic bunch in the interaction region. The ionization pulse follows the excitation pulse by a few tens of nanoseconds, which limits the exposure of the atoms to its strong field during the excitation and so reduces AC Stark shifts~\cite{deGroote2015,DeGroote2017}.

\subsection{Scan control}
A spectrum is recorded by scanning the seed laser continuously across the required frequency range. The two hyperfine branches of a spectrum, the groups of transitions from the $F=4$ and $F=3$ ground levels about 9.2~GHz apart, are recorded as separate acquisitions. Every bunch is logged with the most recent wavemeter reading and acceleration voltage, so each event carries the frequency at which it was recorded, and the spectrum is formed by binning the events in frequency during the analysis. The wavemeter reading was updated every 0.3 to 0.6~s and the voltage about once per second. In the $^2P_{3/2}$ acquisitions only the voltage setpoint was stored. Each branch is swept several times in the course of a measurement, so that slow drifts in the laser frequency or the beam energy average over a branch rather than accumulating across it.
\subsection{Data acquisition}

\setcounter{footnote}{2}Data acquisition is handled by a Python framework developed for this beamline\footnote{\url{https://github.com/ema-lab-mit/capi-daq}}. Pulses from the ion counter are amplified and discriminated, then timestamped against the bunch-release trigger by a Cronologic TimeTagger4-5G time-to-digital converter, which separates pulses only a few nanoseconds apart. At the peak of the resonances, the detector recorded fewer than one ion per bunch, so the acquisition does not saturate, and the recorded count rate is proportional to the ion flux. The resulting pile-up loss is below one percent, and no correction was applied.

Each event is streamed to a time-series database~\cite{influxdb}, so that the spectra can be fitted and monitored in real time and analyzed offline from the same data. The time-of-flight gate applied during acquisition serves only for real-time monitoring. All events are stored regardless, so the gate used in the analysis can be re-optimized without repeating the measurement.

\section{Ion-optical simulations} \label{section:simulation}

Ion-optical simulations of the beamline were performed using the charged-particle trajectory simulation software SIMION~\cite{DAHL20003}. The primary objectives were twofold: first, to determine optimal electrode voltage settings for efficient beam transport along the full beamline; and second, to keep the beam well collimated, so as to maximize the spatial overlap between the atom beam and the laser beam in the interaction region. Reliable ion--laser overlap is critical for high-sensitivity collinear laser spectroscopy measurements, and SIMION simulations have become a standard tool in the design and optimization of such beamlines~\cite{Imgram2023, Imgram2025, Maier2023}.

The electrode geometry of the beamline was first designed using the CAD software Autodesk Inventor. The CAD models were simplified and imported into SIMION, where the potential arrays were generated. The simulated beamline included the following electrodes and grounded boundaries: the exit electrodes of the RFQ cooler-buncher (the RF rods at their DC offset, the exit endcap at its extraction potential, and the exit cone), the ion extractor that accelerates the bunch from the high-voltage platform into the grounded beamline, a set of electrostatic quadrupole triplets for beam focusing and steering, electrostatic benders and deflectors, and the simplified, grounded vacuum fittings connecting the different sections. The radio-frequency field of the trap was not included in the model, nor was the switching of the electrodes from trapping to extraction. The simulation therefore does not describe the trapping and cooling phase. Instead, all electrodes were held at their extraction potentials throughout the simulation.

The ions were launched inside the RFQ cooler-buncher from a Gaussian spatial distribution with a standard deviation of 3~mm in each direction, with directions filling a cone of 45$^\circ$ half-angle about the beam axis. Their kinetic energies followed a Gaussian distribution of mean 30~meV and standard deviation 10~meV, and they were then accelerated to 10~keV by the potential difference between the high-voltage platform and the grounded beamline. These starting conditions are an idealized input rather than a measured property of the beam. The time structure of the bunch was not modeled, so the simulation addresses transverse transport only. A cross-section of the simulated beamline, along with the trajectories followed by the ions and the neutralized species, is plotted in Fig.~\ref{fig:experimental_layout}(b).

The voltages applied to the 17 independent ion-optics elements are strongly correlated, leading to a high-dimensional parameter space that cannot be efficiently explored by manual tuning alone. Moreover, SIMION evaluates ion trajectories through direct numerical integration of the equations of motion on interpolated potential arrays, so no analytical gradient of the transmission with respect to the electrode voltages is available. Thus, a gradient-free optimization strategy is required. In addition, each SIMION evaluation requires approximately 20 minutes of computation time, making the total optimization campaign computationally intensive and further motivating the use of an efficient, sample-aware search algorithm.

To address this, an optimization loop was implemented in which SIMION is driven through Lua~\cite{ierusalimschy1996lua}, and candidate voltage configurations are proposed by the Optuna framework~\cite{optuna_2019}. Each trial launches an ensemble of 1000 ions through the full beamline and returns two figures of merit: the total transmission through a kinematic acceptance window of 19~mm in diameter, set by the clear aperture of the fully opened diagnostic irises that recur along the beamline, and the normalized standard deviation of the transverse ion speed evaluated at the end of the beamline where the ion counter is installed. These two objectives are optimized jointly using the NSGA-II multi-objective evolutionary algorithm~\cite{deb2000fast}, which evolves a population of solutions maximizing the transmission while constraining the beam width in the region of interest. 

\begin{figure}[pos=!ht]
    \centering
    \includegraphics[width=\linewidth]{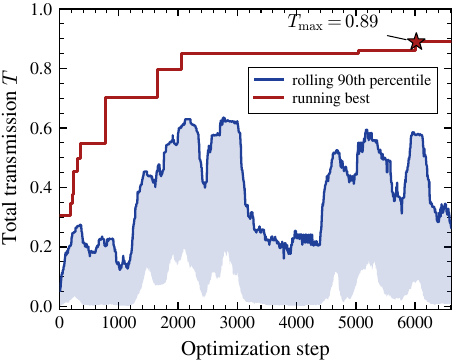}
    \caption{Convergence of the SIMION-Optuna optimization of the 17 ion-optical electrode voltages. The red trace is the running best, reaching $T_{\mathrm{max}} = 0.89$ after ${\sim}\,6$k evaluations. The blue curve marks the rolling 90th percentile, and the shaded band beneath it spans the rolling 50th to 90th percentile over a window of $w = 300$ steps.}
    \label{fig:simion_optimization}
\end{figure}

Figure~\ref{fig:simion_optimization} shows the evolution of the total transmission as a function of the number of optimization steps, reaching a best value of $T_{\mathrm{max}} = 0.89$ after approximately 6\,000 trials. The rolling 90th percentile illustrates the interplay between exploitation and exploration characteristic of evolutionary algorithms: the optimizer alternates between refining high-performing regions and sampling less-explored parts of the parameter space, occasionally discovering improved configurations (e.g., near steps 4\,500 and 6\,000). The voltage set corresponding to $T_{\mathrm{max}}$ was provided as the starting point for the subsequent experimental tuning of the beamline. This is necessary in part due to experimental imperfections, such as misalignment in the beamline elements or machining tolerances, which are not considered in the simulation. In operation, up to 76\% of the ion current measured before the CEC reached the ion counter, compared with 89\% in the simulation, which covers the longer path from the RFQ cooler-buncher to the ion counter.

\section{Experimental results}

The CREMA setup was commissioned using stable $^{133}$Cs. Atoms in the $6s\,^2S_{1/2}$ ground state were excited to the $8p\,^2P_{1/2}$ and $8p\,^2P_{3/2}$ states (rest-frame wavelengths 389.0 and 387.7~nm). The $^2P_{3/2}$ spectra were recorded first, at a beam energy of 10~keV with sodium in the CEC, and the $^2P_{1/2}$ spectra at 15~keV with potassium. The later measurements used the $^2P_{1/2}$ transition because the upper-state hyperfine structure of $^2P_{3/2}$ is not resolved at the present linewidth.

The resonant excitation used the frequency-doubled output of the injection-seeded Ti:Sapphire laser, with pulse energies of 0.2~$\mu$J for $^2P_{1/2}$ and 0.1~$\mu$J for $^2P_{3/2}$, measured at the laser table and kept low to avoid saturating the transitions. About 40\% of this reached the entrance of the beamline. The non-resonant ionization step was provided by the fundamental 1064~nm output of the Quantel Merion MW laser, with pulse energies of 15~mJ for $^2P_{1/2}$ and 20~mJ for $^2P_{3/2}$, overlapping the atomic bunch $\sim$50~ns after the excitation laser pulse.

\begin{figure}[pos=!ht]
    \centering
    \includegraphics[width=\linewidth]{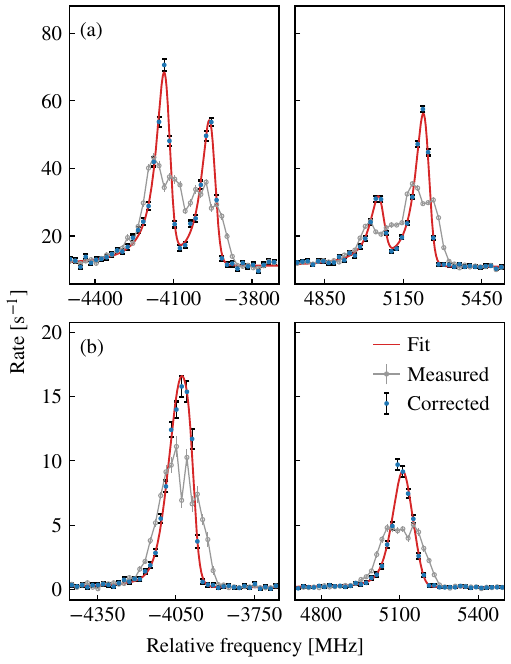}
    \caption{Hyperfine spectra of $^{133}$Cs for the (a) $6s\,^2S_{1/2} \to 8p\,^2P_{1/2}$ (389.0~nm) and (b) $6s\,^2S_{1/2} \to 8p\,^2P_{3/2}$ (387.7~nm) transitions, as measured (gray; the line joining the points only guides the eye) and after the correction of the 60~Hz modulation (blue). Red lines: the fit to the corrected spectra. Error bars are statistical. Frequencies are in the rest frame of the atoms, relative to the fitted center of gravity of each transition. $A_u$ was fixed to its literature value in (b).}
    \label{fig:spectra}
\end{figure}

For the $6s\,^2S_{1/2} \rightarrow 8p\,^2P_{1/2}$ transition (Fig.~\ref{fig:spectra}(a)), two pairs of peaks are observed, corresponding to the allowed $\Delta F = 0, \pm 1$ transitions between the $F = 3, 4$ hyperfine levels of the lower $^2S_{1/2}$ and upper $^2P_{1/2}$ states. Their spacings are set by the magnetic dipole hyperfine constants of the lower and upper states, $A_l$ and $A_u$. A similar structure is observed for the $6s\,^2S_{1/2} \rightarrow 8p\,^2P_{3/2}$ transition (Fig.~\ref{fig:spectra}(b)). Its upper state has four hyperfine levels ($F = 2$ to 5), but with $A_u \approx 7.6$~MHz neighboring components are only 22.8 to 38.0~MHz apart, less than the linewidth of 63~MHz (see below), so they are not resolved. The line shapes were modeled with the SATLAS2 package~\cite{Gins2024}. The electric quadrupole and magnetic octupole constants of $^2P_{3/2}$ were fixed at their literature values~\cite{PhysRevA.107.012807}, far too small to affect the fit.

The spectra were analyzed bunch by bunch. In the $^2P_{1/2}$ data almost half of the recorded pulses are electronic echoes, which follow a real pulse after 3 to 21~ns. They were removed by discarding every pulse that arrives less than 25~ns after the previous one in the same bunch. The $^2P_{3/2}$ data contain none. The spectra were fitted with a Poisson likelihood, comparing the number of ions in each 20~MHz frequency bin with the number expected from the time spent at that frequency. With the corrections described below, the reduced $\chi^2$ of the fits is about 1.1 for $^2P_{1/2}$ and 1.0 for $^2P_{3/2}$. The statistical uncertainties were obtained with a block bootstrap rather than by scaling the fit errors. The data were divided into blocks of 30~s, which were drawn at random with replacement to build resampled data sets, and the fit was repeated on each of them, keeping the phase distribution of each bunch (described below) from the full data set. The same was done with blocks of 60~s, and the larger of the two spreads is quoted.

In both measurements, every hyperfine component is broadened into a flat-topped, double-peaked profile. The cause is a modulation of the beam energy at the 60~Hz mains frequency. The potential of the high-voltage platform, which defines the beam energy, oscillates at this frequency on top of the acceleration voltage. Measured against ground, the oscillation has an amplitude of 4.1~V at 0~kV and 4.7(1)~V at 10~kV, in agreement with the 4.7~V extracted independently from fits to the hyperfine spectra.

The mains phase was not recorded during these measurements, but the modulation can still be removed. The ions of one bunch leave the RFQ cooler-buncher within a few microseconds, a small fraction of the 16.7~ms mains period, so they all see the same shift. The bunches are released every 10~ms, or every 40~ms in the $^2P_{3/2}$ measurement, so the modulation advances by 0.6 or 0.4 of a cycle from one bunch to the next, while the mains frequency drift is slower. A single bunch gives less than one detected ion on average, but the counts of the neighboring bunches over a few seconds determine the phase of the mains cycle at which each bunch was released. This phase was estimated with a hidden Markov model~\cite{Rabiner1989}, in which the phase advances by a fixed step plus a slow random drift and the counts of each bunch follow Poisson statistics around the shifted spectrum. The phase of each bunch was estimated from the counts of the other bunches only, so that its own counts cannot bias it. Each bunch was then given the rest-frame frequency of the Doppler correction with the recorded voltage plus the estimated modulation at its phase, with its counts unchanged. The fit averages the line shape of each bunch over the remaining uncertainty of its phase. After the correction, the fitted lines are 55~MHz ($^2P_{1/2}$) and 63~MHz ($^2P_{3/2}$) wide (FWHM).

Even after the correction, fits with symmetric Voigt profiles leave residuals that are systematically positive on one flank of every hyperfine component and negative on the other, with reduced $\chi^2$ values of 11.0 for $^2P_{1/2}$ and 2.4 for $^2P_{3/2}$, so the observed line profile is asymmetric. A plausible origin is a tail of the velocity distribution of the neutralized atoms toward low velocities, to which the charge exchange can contribute~\cite{DOCKERY2025}. In the co-propagating geometry used here, slower atoms appear at lower frequency, the side on which the fit places the tail. The spectra were therefore fitted with a skewed Voigt profile, using a single asymmetry parameter shared by all components. With this profile, the reduced $\chi^2$ drops to the values quoted above.

For $^2P_{1/2}$, where the hyperfine components are resolved, the relative intensities were left free. For $^2P_{3/2}$, the components of each branch lie closer together than the linewidth, where a change in their relative intensities has nearly the same effect on the profile as a change in the splitting or in the width. The relative intensities within each branch were therefore fixed to the Racah intensities.

The fitted lines (55 and 63~MHz FWHM) are more than twice as wide as the laser contribution of about 21~MHz expected after frequency doubling~\cite{sonnenschein2015thesis}. The additional width may come from a spread in the kinetic energy of the atoms, to which the extraction from the RFQ cooler-buncher and the shape of its trapping potential, space charge in the ion bunches, collisions with residual gas during acceleration, and the charge exchange in the CEC can all contribute. The excitation itself may also add width, since the low pulse energies reduce saturation but may not remove it completely. These contributions cannot be separated with the present data. 

The ion counter records about 10 to 20 counts per second that are not correlated with the bunches, such as its own dark counts. The time-of-flight gate is open for only 3 to 4~$\mu$s, so these counts add less than 0.01 per second to the spectra. The background that remains within the gate arrives with the bunches, at a rate of about 11 ions per second for $^2P_{1/2}$ and 0.2 ions per second for $^2P_{3/2}$, a sixth and a hundredth, respectively, of the tallest peak of the corrected spectra. Per bunch, the tallest peaks of the two spectra are about equally high, so the background per bunch is about fifteen times higher in the $^2P_{1/2}$ measurement, which differs from the $^2P_{3/2}$ one in alkali vapor in the CEC, beam energy, transition, and laser power. In the $^2P_{3/2}$ measurement, the ionization laser pulses fired while no bunch was in the interaction region produced no counts above the dark-count rate. The background is therefore related to the atom beam and is not caused by the laser pulses alone. It may arise from ionization of the neutral atoms in collisions with residual gas, or from their non-resonant ionization by the laser light.

\subsection{Hyperfine parameter extraction}
\begin{table}
\caption{Magnetic dipole hyperfine constants of the lower ($A_l$) and upper ($A_u$) states from the commissioning spectra of $^{133}$Cs, compared with literature values from Refs.~\cite{PhysRevA.107.012807,BIPM2019SI}. Statistical and systematic uncertainties are given in parentheses $(\,)$ and square brackets $[\,]$, respectively.}
\centering
\begin{threeparttable}
\renewcommand{\arraystretch}{1.3}
\begin{tabular}{l r@{.}l r@{.}l}
\hline\hline
 & \multicolumn{2}{c}{This work (MHz)} & \multicolumn{2}{c}{Literature (MHz)} \\
\hline
\multicolumn{5}{l}{$6s\,^2S_{1/2}\to 8p\,^2P_{1/2}$} \\
\quad $A_l$ & 2297 & 3(3)[8]  & 2298 & 1579425\tnote{a} \\
\quad $A_u$ &   43 & 2(3)[1]  &   42 & 933(8) \\
\multicolumn{5}{l}{$6s\,^2S_{1/2}\to 8p\,^2P_{3/2}$} \\
\quad $A_l$ & 2299 & 3(4)[28] & 2298 & 1579425\tnote{a} \\
\quad $A_u$ & \multicolumn{2}{c}{fixed} & 7 & 609(8) \\
\hline\hline
\end{tabular}
\begin{tablenotes}
\item[a] Exact by definition. The $^{133}$Cs ground-state hyperfine splitting is fixed at $\Delta\nu_{\mathrm{Cs}} = 9\,192\,631\,770$~Hz by the SI definition of the second~\cite{BIPM2019SI}, resulting in $A_l = \Delta\nu_{\mathrm{Cs}}/4$ with no associated uncertainty.
\end{tablenotes}
\end{threeparttable}
\label{tab:hyperfine_parameters}
\end{table}

The extracted hyperfine constants are summarized in Table~\ref{tab:hyperfine_parameters}. Both determinations of the lower-state constant agree with the value fixed by the SI definition of the second, within 0.9 ($^2P_{1/2}$) and 0.4 ($^2P_{3/2}$) standard deviations of their total uncertainties, and $A_u$ of $^2P_{1/2}$, left free in the fit, agrees with the literature value~\cite{PhysRevA.107.012807} within one standard deviation.

For $^2P_{3/2}$, the upper-state components are not resolved, so $A_u$ was fixed to its literature value of 7.609(8)~MHz~\cite{PhysRevA.107.012807} and only $A_l$ was fitted. Varying $A_u$ within its quoted uncertainty changes $A_l$ by about 10~kHz. Leaving $A_u$ free instead results in $A_u = 9.0(3)$~MHz, well above the literature value, and moves $A_l$ by 2.4~MHz. With the components this close together, a free $A_u$ absorbs imperfections of the line shape. The shift of $A_l$ is counted as a model uncertainty.

\subsection{Systematic uncertainties}\label{subsection:systematics}

(i) \textit{Voltage drifts}. In the $^2P_{1/2}$ acquisitions, the acceleration voltage stored with every event was used in the Doppler correction, so its drift is corrected rather than treated as an uncertainty. The voltage drifts at about 0.45~V/h, and the mean voltages of the two hyperfine branches of a spectrum differ by 0.33~V. Left uncorrected, this would shift the measured ground-state splitting by 4.1~MHz, or $A_l$ by 1.0~MHz. The scatter of the individual readings, 0.5~MHz on the frequency axis, averages down to below 10~kHz on $A_l$. For $^2P_{3/2}$, where only the setpoint was stored, a drift of 0.3~V between the two branches, comparable to that measured for $^2P_{1/2}$, is counted as a systematic uncertainty of 1.2~MHz on $A_l$.
    
(ii) \textit{Frequency measurement.} The manufacturer-specified absolute accuracy of the HighFinesse WS8-2 wavemeter is 2~MHz at the calibration wavelength of a HeNe laser and at most 10~MHz at the operating wavelength around 775~nm. The 10~MHz reflects an offset that varies only slowly with wavelength~\cite{Knig2020}, which shifts all measured frequencies by nearly the same amount and cancels in the frequency differences that determine the hyperfine constants. The remaining 2~MHz contributes 0.8~MHz to $A_l$.

(iii) \textit{Analysis.} Several sources of systematic uncertainty were considered in the extraction of the hyperfine parameters. Varying the widths of the frequency bins used in the fits changed the extracted value of \(A_l\) for \(^{2}P_{1/2}\) by up to 0.1~MHz. For \(^{2}P_{3/2}\), fixing \(A_u\) adds the model uncertainty of 2.4~MHz on \(A_l\) described above.

We further assessed the sensitivity of the results to corrections in the fitting procedure and to the assumed line-shape model. These tests yield uncertainties of 0.04~MHz for \(A_l\) and 0.07~MHz for \(A_u\) of the \(^{2}P_{1/2}\) state, and 0.15~MHz for \(A_l\) of the \(^{2}P_{3/2}\) state. All these contributions are included in the final uncertainty budget.

\section{Conclusions and outlook}\label{sec:conclusions}
The CREMA collinear resonance-ionization spectroscopy setup has been constructed and commissioned at the MIT EMA Laboratory. High-resolution hyperfine spectra of the \(6s\,^2S_{1/2} \rightarrow 8p\,^2P_{1/2}\) and \(6s\,^2S_{1/2} \rightarrow 8p\,^2P_{3/2}\) transitions in \({}^{133}\mathrm{Cs}\) were recorded, and the extracted hyperfine constants agree with literature values within the combined uncertainties. These results demonstrate the readiness of CREMA for precision laser-spectroscopy measurements.

The next upgrades will further improve the precision and versatility of the setup. Completion of the charge-exchange-cell voltage scanner will allow Doppler tuning of the atomic beam, while an iodine-stabilized HeNe reference will improve the absolute calibration and long-term stability of the wavemeter.  A quantitative measurement of the overall resonance-ionization efficiency will also be required to assess how ionization schemes developed at CREMA can be transferred to low-intensity online radioactive-beam experiments.

As an offline counterpart of the RISE experiment at FRIB, CREMA provides a platform for developing and validating ionization schemes, characterizing systematic effects, and testing new spectroscopy methods before scarce radioactive-beam time is used. It will therefore support future precision studies of exotic atoms and molecules, including measurements for nuclear structure, astrophysics, and fundamental symmetries.

\section*{Declaration of competing interest}
The authors declare that they have no known competing financial interests or personal relationships that could have appeared to influence the work reported in this paper.
\section*{Data availability}
The event data and the analysis code that support the findings of this study are openly available in Zenodo~\cite{CREMAdata2026}.
\section*{Declaration of generative AI and AI-assisted technologies in the manuscript preparation process}
During the preparation of this work, the authors used Claude (Anthropic) as an assistant for writing, for developing the analysis code, for checking the references, and for proofreading. After using this tool, the authors reviewed and edited the content as needed and take full responsibility for the content of the published article.

\section*{Acknowledgements}
We acknowledge support from the Jarve Seed Fund for Science Innovation at MIT. This work was in part supported by the U.S. Department of Energy, Office of Science, under grant DE-SC0021176.
F.~C.~P.~C. acknowledges support from the Colciencias scholarship of Colombia and the Santo Domingo Fellowship at MIT. J.~K. acknowledges the support of a Feodor Lynen Fellowship of the Alexander-von-Humboldt Foundation.
We are grateful for the technical support provided by the Bates Laboratory at MIT. In particular, we thank Ernest E. Ihloff, Chris Vidal, John McGlashing, and Joseph D. Cucinotta for their contributions to the technical design and implementation of the setup. We acknowledge Xiaofei Yang for generously providing the charge-exchange-cell (CEC) chamber.
The design of the RFQ cooler-buncher received funding from the European Research Council (ERC) under the European Union's Horizon 2020 research and innovation programme under grant agreement No. 679038. M.~R. acknowledges support from the Research Council of Finland via Project No.~368322 (NUANZ).

\bibliographystyle{elsarticle-num}
\bibliography{Bibliography}

\end{document}